\documentclass[11pt]{article}

\usepackage{mathptmx}

\usepackage[latin9]{inputenc}

\usepackage{color}
\definecolor{document_fontcolor}{rgb}{0, 0, 0}
\color{document_fontcolor}

\usepackage{pdflscape}

\usepackage{tabularx}
\usepackage{authblk}
\usepackage[numbers,sort]{natbib}

\makeatletter

\author[1,2,3,4]{Florian Klimm\thanks{Both authors contributed equally.}}
\author[5,6]{Benjamin F. Maier$^*$}
\affil[1]{University of Oxford, Oxford OX2 6GG, United Kingdom}
\affil[2]{Department of Statistics, University of Oxford, 24-29 St Giles', Oxford OX1 3LB, United Kingdom}
\affil[3]{Department of Mathematics, Imperial College London, London, SW7 2AZ, United Kingdom}
\affil[4]{MRC Mitochondrial Biology Unit, University of Cambridge, Cambridge Biomedical Campus Hills Road, Cambridge, CB2 0XY, United Kingdom}
\affil[5]{Department of Physics, Humboldt Universit\"at zu Berlin, Newtonstra\ss{}e~15, D-12489 Berlin, Germany}
\affil[6]{Robert Koch-Institute, Nordufer 20, D-13353 Berlin, Germany}

\date{\today}

\usepackage{lipsum}

\usepackage{url}
\usepackage{listings}
\usepackage[dvipsnames]{xcolor}
\usepackage{rotating}

\title{A Network Science Summer Course for High School Students}

\colorlet{maincolor}{black}

\usepackage{listings}

\makeatother

\usepackage{listings}

\begin{document}
\maketitle 
\begin{abstract}
We discuss a two-week summer course on Network Science that we taught
for high school pupils. We present the concepts and contents of the
course, evaluate them, and make the course material available.

\vspace{1cm}

\end{abstract}

\section{Introduction \& Motivation}

Teaching \emph{network science}~\cite{newman2010networks} has been
identified as an important endeavor, because it is a concept usually omitted in high
school curricula~\cite{cramer2015netsci,network_literacy,sayama2015essential}.
Hence, scientists organized outreach events for teenagers to bring
them into contact with the study of networks, notably in England~\cite{harrington2013commentary}
and Spain~\cite{sanchez2014more}. These outreach events consisted
of an introductory presentation, followed by work in smaller groups
on topics such as `node importance', `disease spread and vaccination
strategies', and `why your friends have more friends than you do'.
Based on the success of these half-day events we decided to design
a two-week summer course for German high school pupils with the topic
`Networks and Complex Systems' (German: \emph{Netzwerke und komplexe
Systeme}). The course was part of an established German summer school
called \emph{Deutsche Sch\"ulerakademie} (German Pupils Academy), whose
concepts will be introduced in Section~\ref{sec:DSA}.

Due to the longer duration of the course and the higher age of the
participants we were allowed to extend the covered material from the
shorter outreach events, go into more mathematical detail with the
topics, see Section~\ref{sec:contents}, and include some components
of programming with Python. In Section~\ref{sec:conclusions} we
discuss our approaches and evaluate to what extent they were successful
or might need further improvements. We make our teaching material
available online~\cite{github}.

\section{Concept of the \emph{Deutsche Schülerakademie}}

\label{sec:DSA}

The \emph{Deutsche Schülerakademie} (DSA) is an extracurricular summer
school for motivated and gifted pupils in the final two years of German
high-school~\cite{DSA}. For more than twenty years every summer
multiple of these academies have been organized where each of them
consists of six courses for 15\textendash 16 pupils each. Our course
was part of the academy `Roßleben 2016' and set in a monastery school in a small town in Thuringia. The courses concurrent with
ours covered diverse academic topics from analyzing Richard Wagner's
operas to String Theory. Approximately one hundred pupils participated
in the courses and were encouraged to organize academic and non-academic
extracurricular activities. Over the two weeks the participants spent
a total of approximately sixty hours in their courses. In this time
they worked under guidance of the lectures. Furthermore, they had
to prepare talks for the participants of other courses explaining
the topics of their course, a concept called \emph{Rotation} (see
Subsection~\ref{subsec:rotation}), and create a written summary
of the learned material, called \emph{Dokumentation} (see Subsection~\ref{subsec:doku}).

\section{Course Contents \& Didactical Elements}

\label{sec:contents}

In the course we used different didactic elements. Some of them, as
the \emph{Rotation} and the \emph{Dokumentation}, were framed by the
DSA structure, others, as pupil presentations and programming exercises,
were chosen deliberately by us because they seemed appropriate for
conveying the material.

\subsection{Participant Questionnaire}

We received a list with contact details of all fifteen participants
two months before the academy. We expected some heterogeneity in existing
knowledge because the pupils were in one of the final two grades in
German high school. This discrepancy is enhanced by the federal education
system in Germany which gives different foci to subjects in different
states~\cite{lohmar2014education}. To assess the knowledge of the
pupils we conducted a non-anonymous online survey which was answered
by all fifteen participants. We mainly assessed the mathematical knowledge
in eleven brief questions, for example, `Do you know what a matrix
is?'. The results gave us an overview over the pupils' abilities:
All pupils knew how to operate with vectors and two thirds knew how
to work with matrices, almost all of them were able to differentiate,
but less knew about Riemann integration. The most heterogeneous knowledge
was given for programming with a third of the participants having
no programming experience and others knowing Python or C-family programming
languages rudimentarily. Three quarters did not know \LaTeX{}, which
we intended to use for the \textit{Dokumentation}. We also included an open-ended question about their motivation to choose this particular course. The pupils stated that they were especially interested because the course connects mathematical tools with real-world applications and targets questions from different sciences as physics and biology.

\subsection{Reader}

Based on the results from the questionnaire we provided the participants
with a so-called \emph{Reader} as preparation for the course. The
main purposes of this document was to give all participants a common
level of mathematical background and to provide them with some programming
experience. For this we drew from two sources, (i) the Python online
course `Spielend programmieren lernen' (engl. `Learning Programming
Effortlessly') by the Hasso-Plattner-Institute~\cite{meinel2017openhpi,spielend_programmieren}
and one chapter of `Physik mit Bleistift' (engl. `Physics with a Pencil')
~\cite{schulz2004physik}. The latter gave a brief introduction
to vectors and differentiating. In the post-course evaluation the
pupils were, however, not satisfied with the material as it focused
too much on mathematical tools that we used only briefly and not enough
on learning programming.

\subsection{Pupil Presentations}

Each of the participants gave a presentation during the course. In
preparation of the course we suggested a topic to each of them based
on their interests as deduced from the survey and provided them with
reading material. The talks were supposed to be 10\textendash 15\,min
long and we advised the participants to approach the topics from a
phenomenological side. For example, Turing patterns were introduced under
the topic `how leopards get their patterns'. Our aim was to discuss
the mathematics behind the phenomena together after the introduction
given by the pupils. An overview over all fifteen topics and a brief
assessment of their suitability is shown in Table~\ref{tab:talks}.
We noted that some pupils considerably overrun the allocated time.
We decided to not interrupt their talks but let them know afterwards
that this might not be appropriate in other circumstances.

\begin{table}
 \caption{Overview of the topics presented by pupils in talks. }
  \label{sample-table}

    \begin{tabularx}{\linewidth}{r@{\hskip0.1cm} | @{\hskip0.1cm}X}
      \hline\hline
	Topic &Comment  \\ \hline
Four color map problem  & 
Good introduction to graph coloring. Allows a discussion about computer-assisted mathematical proofs \\

Seven bridges of K\"onigsberg  & Classic and easily accessible problem \\

Traveling salesman problem  & Allows introduction to computational complexity \\
Kirchhoff's circuit laws  &  Well-suited to introduce pupils without a physics background to physical phenomena\\

Percolation theory  & Easily accessible with e.g. forest tree density and forest fires \\
Cybernetics  & Participant introduced a school project's work \\
Neuronal networks  & Interesting topic but too complex for deep discussion \\
Graph isomorphism problem  & Allows discussion of $\mathcal{P}$ vs. $\mathcal{NP}$ \\

Disease spreading & Excellent for tying together network phenomena and dynamics \\
Predator-prey interactions  & Good introduction to dynamical systems but too complex for deep discussion \\

Turing patterns  & Accessible with, e.g., animal fur patterns but too complex for deep discussion \\
Deterministic chaos  & The logistic map is well-suited for simple analyses of chaotic phenomena \\

Fractals  & Nicely accessible with, e.g. coast lines and Koch's curve \\

Fractal dimension  &  Counter-intuitive but easily accessible with box counting method \\ \hline
    \end{tabularx}

\label{tab:talks} 
\end{table}

\subsection{Lectures}

Lectures were a fundamental part of our course. Usually, they covered
concepts that were earlier introduced by the pupils in their presentations.
We subsequently formalized the mathematical description and went into
more detail. We briefly summarize the presented topics below.

\subsubsection{Introduction to Graph Theory}

This lecture was built on the discussion about the representation
of graphs as edge lists and adjacency matrices (see Subsubsection~\ref{sssec:graph}).
We proceeded to introduce simple network measures as the number of
edges $m$, node degree $k$, path length $L$ and density $\rho$.
We then discussed some deterministic synthetic networks as the empty
graph $N_{n}$, the path graph $P_{n}$, the complete graph $K_{n}$,
and the complete bipartite graph $K_{n,n}$ and derived formulas of
the network measures in dependence of the graph size $n$. Later in
a programming exercise the student validated these formulas numerically.

\subsubsection{Graph Coloring}

Our aim was to teach the pupils network science and classical graph
theoretical concepts alike. \emph{Graph coloring} was an appropriate
field for the latter because the underlying problems are easily accessible,
while the proofs can be challenging. We introduced node coloring
and edge coloring but focussed on the former. The pupils particularly
enjoyed proving theorems and challenged our proofs if they were too
sloppy. In the last part of this lecture we also discussed coloring
algorithms, particularly \emph{greedy coloring}. One task the pupils
enjoyed was to compute an upper bound of the chromatic number $\chi$
of a symmetrize food web~\cite{ulanowicz2005network} consisting
of $n=63$ animals in the Everglades by applying greedy coloring
many times to different orders of the nodes. We motivated this question
by asking how many different cages are necessary if a zoo director
would like to keep all the animals without one preying on the other.

\subsubsection{Social Networks \& Friendship Paradox}

The discussion of social networks is well-suited to motivate the real-world
applicability of network science to the pupils. The participants were
rather astonished by the so-called \emph{friendship paradox}. We followed
Strogatz's excellent explanation in the New York Times~\cite{friendship_paradox}.

We proceeded by discussing the existence of clusters and other mesoscale
structures in networks. Unfortunately, for privacy reasons, Facebook
does not allow downloading your own friendship network anymore. This
used to be an excellent way to make pupils familiar with various
network characteristics as communities and centrality measures. Therefore,
we had to fall back on other available data, e.g. the social graphs
from \textsc{MovieGalaxies}~\cite{kaminski2012moviegalaxies} and
B. F. Maier's Facebook friends network from 2014 ~\cite{maier_cover_2017}.
As another example of a small-world network we played the `Wikipedia-Game'
to demonstrate short link paths between apparently unrelated articles.
Its objective is to find the shortest link-path between two random
Wikipedia articles by only following links on the encountered Wikipedia
pages~\cite{west2009wikispeedia}.

\subsubsection{Effective Resistance}

A lot of the pupils showed interest in physics. Therefore we dedicated
a day to the discussion of resistor networks. This topic was first
introduced by a pupil's presentation on Kirchhoff's circuit laws.
Subsequently we aimed at computing the effective resistance of more
complex networks where we followed the discussion in~\cite{newman2010networks}.
In particular, we proceeded as follows (see also Exercise Sheet~3).

At first we motivated eigenvalues and -vectors of matrices with a
rudimentary example of a growing population of linearly interacting
species. Using this concept we were able to motivate the inverse matrix
and what it means when a matrix has eigenvalue zero (the concept of
singularity), which was necessary to explain the reduction of the
graph Laplacian to make it invertible.

This topic was very difficult for some of the younger pupils which
were not as versatile in working with matrices as the older ones.
However, some of the more engaged participants particularly enjoyed
those discussions.

\subsubsection{Dynamic Systems and Fractals}

In addition to traditional network science we also discussed \emph{dynamic
systems}, e.g. simple exponential growth or more complex Lotka-Volterra
equations. The pupils enjoyed the discussion, although an introduction
of mathematical prerequisites, e.g. ordinary differential equations
(ODEs), was necessary. Even though introducing those prerequisites
took a rather long time, we think it was valuable for the pupils to
learn about ODEs using very simple systems. In particular we showed
how to derive the equation of motion for logistic growth given the
example of the mean-field model of the susceptible-infected infection
model. Here, we started with discrete time and probabilistic considerations
and arrived at the final equation by going to continuous time using
$\Delta t\rightarrow0$.

We furthermore discussed chaotic systems by analyzing the logistic
map in discrete time, as well as the stability of equilibria since
those are very accessible with high school-level calculus and graphical
methods equally. Especially the discussion of stability was hence
well-received and the participants seemed to grasp the concept of
phase transitions easily by analyzing the logistic growth equation
in continuous time.

The lectures on dynamic systems ended with the discussion of fractals
(e.g., as coast lines) and fractal dimensions (by introducing the
box counting-method). While the pupils seemed to understand the fundamental
concepts they had significant problems reproducing self-similar structures
using Lindenmayer systems and Python's implementation of `turtle',
a drawing engine that uses simple iterative instructions.

In retrospect, we feel that this part of the course was a bit out of
scope and potentially overwhelming for the pupils.

\subsection{Modules}

In addition to longer lectures we also used a couple of smaller modules,
which had a stronger component of pupil involvement. We will briefly
introduce some of them below.

\subsubsection{How to Communicate a Graph}

\label{sssec:graph} 
This module was one of the first elements of
our course. We separated the pupils into two groups. Each of them
was given a printout of a small labeled graph with the aim to communicate
the graph in written form to the other group such that they were able
to recreate the same graph.

The pupils quickly identified ways to achieve this, for example, as
an edge list. After joining both groups together we discussed different
approaches and their limitations, for example, that edge lists do
not allow the representation of isolated nodes but an additional node
set is required. This naturally led to the introduction of other representations
such as adjacency matrices. One of the graphs was dense and
the other sparse, which lead to the discussion of space and memory
considerations when working with large graphs in different representations.

This simple task was very successful because it clearly motivates
the definition of a graph as node and edge set. Giving
the mathematical definitions exclusively can intimidate pupils who are not familiar
with the notion of sets.

\subsubsection{Analog Creation of Random Graphs with Dice}

\begin{figure}[t!]
\centering \includegraphics[width=1\linewidth]{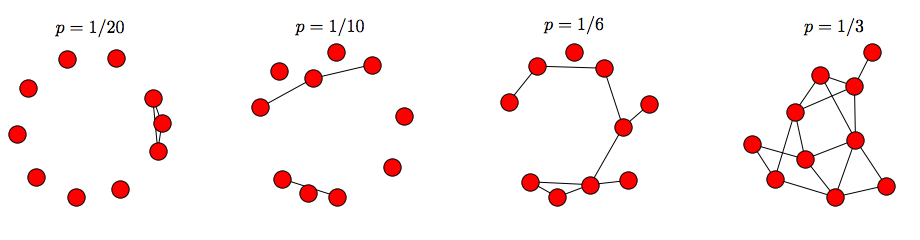}
\caption{Four $G(n=10,p)$ random graphs as generated by the participants by rolling dice. A twenty-sided dice was used for the probabilities $p=1/20$ and $p=10$ and a six-sided dice for $p=1/6$ and $p=1/3$.}
\label{fig:random_graphs} 
\end{figure}

After introducing Erd\H{o}s-Rényi random graphs $G(n,p)$ we gave
the participants the task to create undirected random graphs with
the help of six-sided and twenty-sided dice. We discussed how
to generate graphs with different connection probabilities $p$ and
also identified an appropriate number $n$ of nodes. We settled for
$n=10$ nodes given that $45$ dice rolls per network seemed a reasonable
time effort. We then split the pupils into four groups, each creating a graph
of different connection probability $p \in \{1/20,1/10,1/6,1/3\}$. Each group drew their final graph
on the blackboard (see Fig.~\ref{fig:random_graphs}) and we subsequently
discussed their properties, e.g.~ the size of their giant connected
component and their degree distributions $P(k)$, and compared them with theoretical
expectations.

This task was praised as `very fun' by the pupils. We evaluated it as successful from a didactical
perspective because it playfully connects
known material, such as dice rolls and probabilities, with new concepts, such
as random graphs. Furthermore, it allows the transition to the creation
of random graphs using pseudorandom number generators of a computer.
However, the introduction of the binomial coefficient to explain the
random graph's degree distribution was rather challenging for the
majority of the pupils.

\subsubsection{Creating a Floor Plan Network}

\begin{figure}[t!]
\centering \includegraphics[width=0.4\textwidth]{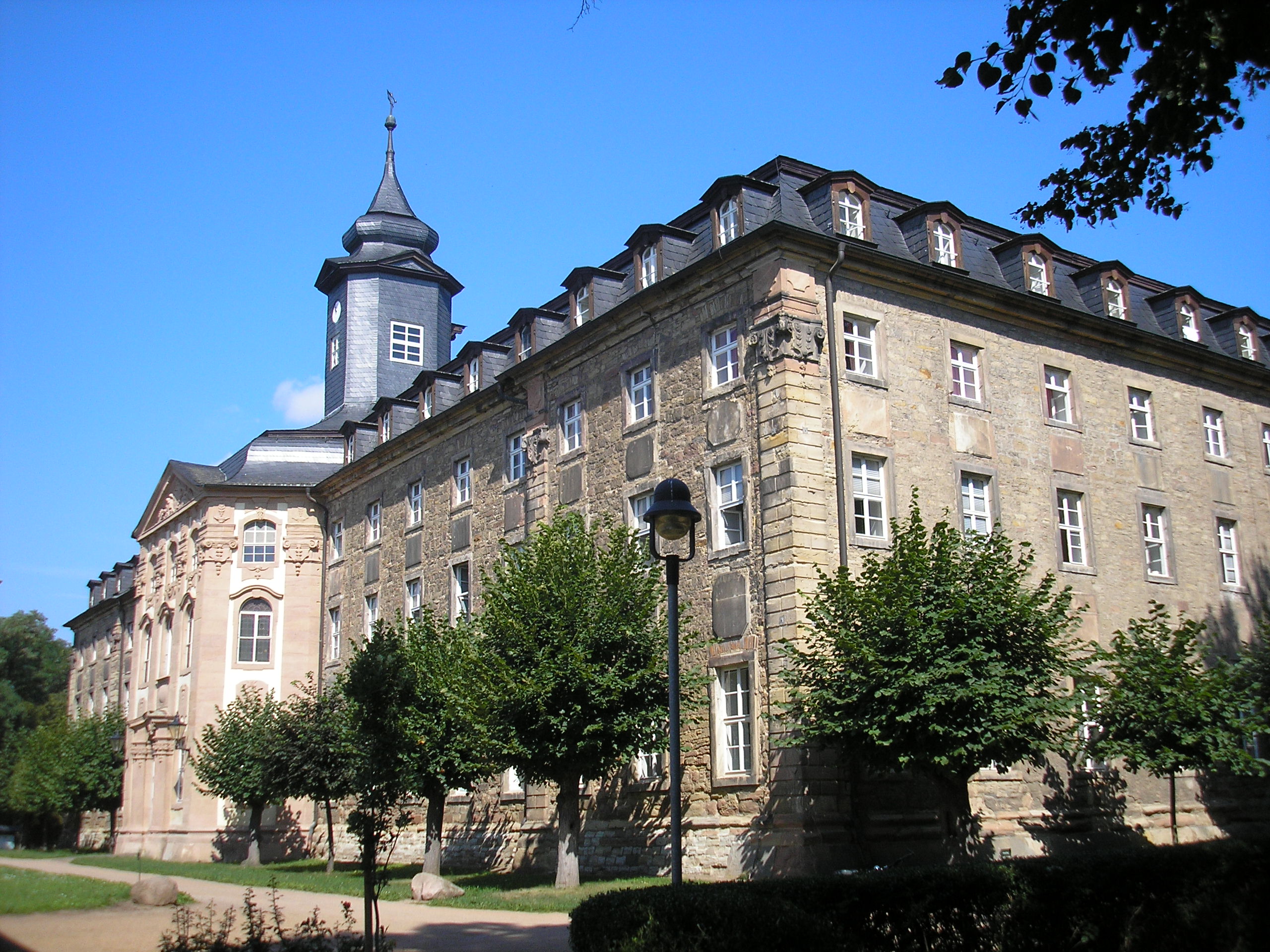}
\includegraphics[width=0.59\linewidth]{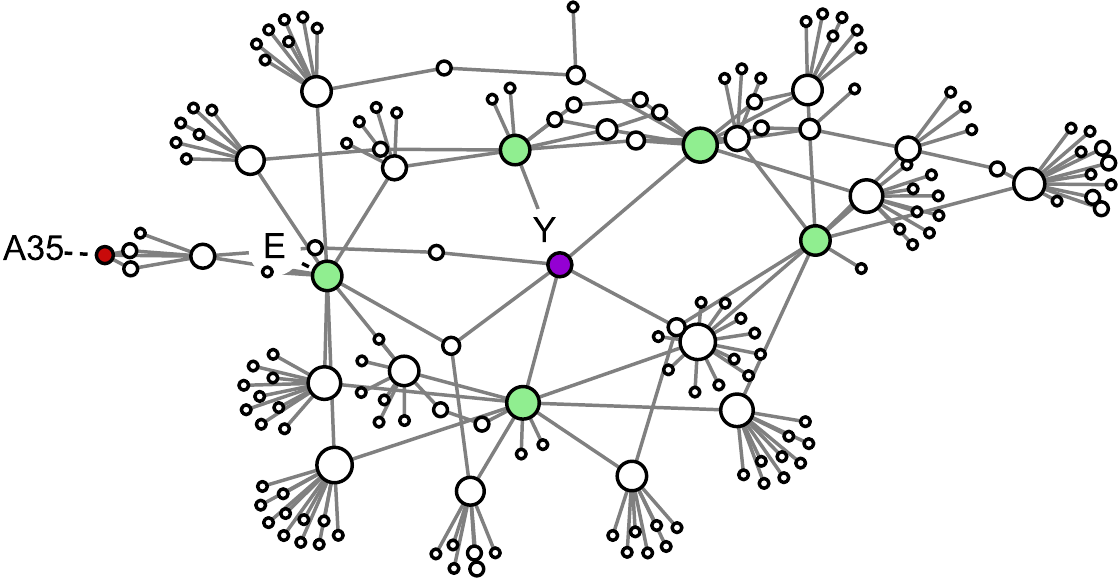}

\caption{Left: The monastery school building in Ro\ss{}leben~\protect\cite{klosterschule_photo}.
Right: Network illustration of the floor plan of the former monastery
school as created by the pupils. Marked are the eastern staircase
`E' which has a high betweenness centrality and is close to the 
epidemic outbreak room `A35', as well as the yard `Y'.}
\label{fig:klosterschule} 
\end{figure}

In addition to synthetic networks and the networks created from data that
we provided, we wanted the participants to experience a more realistic
experience of network creation. For this we gave them access to a
printout of the floor plan of the \emph{Klosterschule Roßleben}, which
hosted the academy (see Fig.~\ref{fig:klosterschule}). They had
the task to create a network in which nodes represent the $n=161$
rooms and edges stand for direct connections between them.

We let the pupils organize amongst themselves how to achieve this
and how to overcome the problems they faced. They decided to split
into groups, each of them covering one floor of the building. They
labeled each room with a unique alphanumerical identifier consisting
of a letter for the floor and a number for the room. The four stairwells,
which connected floors, were given the label `East', `South', `West',
and `North'. The unique labeling allowed the parallel creation of
different parts of the network which were joined together later on.
To avoid making mistakes in the network creation process each
floor's network was generated independently twice. The pupils then used
their knowledge about necessary conditions of graph isomorphism, e.g.
identical number of edges and degree distributions, to check whether both were potentially
identical. If this was not the case they knew that there was a discrepancy
between the created networks and they investigated which of the two creators made
a mistake.

The generated network, which is given in the course material \cite{github}
and shown in Fig.~\ref{fig:klosterschule}, was then further analyzed
by the pupils. For example, they investigated the degree distribution
and compared different centrality measures. They found a positive
correlation between degree and betweenness while some nodes contradicted
that notion by having a much higher betweenness than expected from
their degree. They identified these nodes to represent staircases
that connect the hallways with each other.

\subsubsection{Infectious Dynamics}

\begin{figure}[t!]
\centering \includegraphics[width=0.49\linewidth]{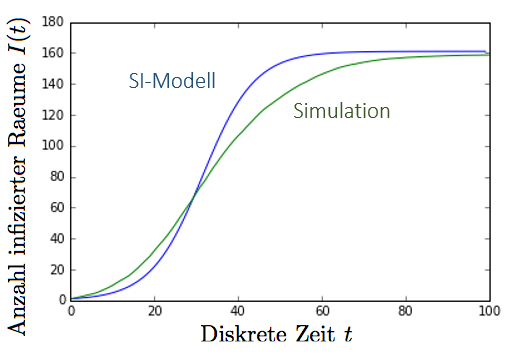}
\includegraphics[width=0.49\linewidth]{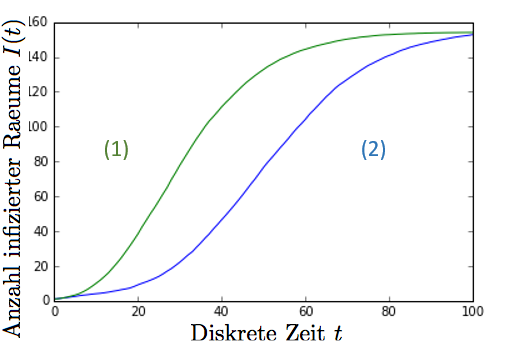}
\caption{Left: Simulated SI dynamics on the monastery room network (green)
in comparison with a mean-field approximation (blue). Right: Spread
of the disease as number of infected rooms over time for random
vaccination of one room
(green, (1)) and targeted vaccination of staircase `E' (blue, (2)).
Both figures were produced by one of the participants and thus have German labels. Horizontal axes give the time steps and vertical axes show the number $I(t)$ of infected rooms.}
\label{fig:SI_dynamics}
\end{figure}

We used the created school building network
for the simulation of compartmental models in epidemiology, specifically
the susceptible-infected dynamics (SI), as shown in Figure~\ref{fig:SI_dynamics}.
We invented a scenario where another course of the academy, working
with bacteria germs, let a hyperinfective strain escape which slowly
infects the whole building room by room. The pupils then compared
the results with the mean-field approximation introduced in the lectures.

We then discussed three vaccination strategies: (i) random vaccination,
(ii) vaccination of nodes with high betweenness, and (iii) random
next-neighbor vaccination. The temporal development of the number
$I(t)$ of infected nodes for the former two are shown in Figure~\ref{fig:SI_dynamics}
and the pupils were able to identify that the betweenness-vaccination
is slowing the spread of the disease. The final steady state, however,
shows the same number of infected rooms.

This module successfully connected different aspects of the course,
betweenness centrality and SI dynamics, with a real-world application.
Furthermore, it demonstrated the effectiveness of vaccination to the
pupils and we were able to discuss the aspect of herd immunity.

\subsubsection{Small-World Networks}

Besides the introduction of essential concepts for networks we also
wanted to familiarize the pupils with scientific literature. For this
we chose the first half of Watts' and Strogatz' famous paper \emph{Collective
Dynamics of `small-world' networks}~\cite{watts1998collective}.
The task we gave them consisted of two parts: (i) read the paper and
explain the definition of \emph{clustering coefficient} and (ii) recreate
Figure~2, which gives the characteristic path length and clustering
coefficient of a lattice graph under rewiring.

\begin{figure}[t!]
\centering \includegraphics[width=0.7\linewidth]{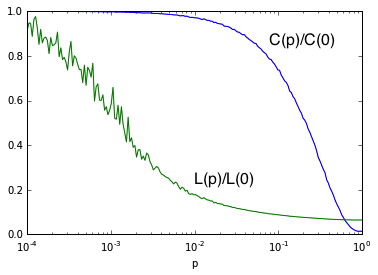}

\caption{Normalized characteristic path length $L(p)/L(0)$ and normalized clustering coefficient $C(p)/C(0)$
for the family of randomly rewired graphs with rewiring probability $p$, as discussed by Watts and
Strogatz~\protect\cite{watts1998collective}. This figure was produced
by a pupil.}
\label{fig:watts_strogatz} 
\end{figure}

This task was challenging but fruitful for the pupils. All of them
quickly understood the contents of the paper. For the programming
exercises they needed more help. We were, however, rather impressed
by the pupils' abilities. Reproducing the results of an important
paper, as shown in Figure~\ref{fig:watts_strogatz}, led to a feeling
of achievement as well as motivation for the upcoming lectures. Note
that for time limitations we omitted averaging over a large number
of realizations, which was used in the original paper. However, we discussed with the students the impact of and reason for such averaging procedures.

\subsection{Problem Sheets}

We created six problem sheets that gave the pupils the chance to apply
some of the learned concepts. We initially planned a problem sheet
per day but quickly realized that our schedule was too ambitious.
The problem sheets usually consisted of a mixture of pen-and-paper
analytical questions, as well as programming exercises as outlined
in the next Subsection. We have made all problem sheets in the original
German and a translated English version available online \cite{github}.

\subsection{Programming Exercises}

We used Python as the main and only programming language because it
is open source and easy to learn and implement. In addition to the
default Python functions we used mainly three libraries, \textsc{networkX},
\textsc{matplotlib}, and \textsc{numpy}. We also used Gephi
\cite{gephi}
for the analysis of social networks and as an easy way to graphically
access them.

Given the results from our initial questionnaire we were aware that
the pupils had very different levels of knowledge of programming. To tackle this
problem we created two groups, one working on their own on the programming
exercises, the other receiving closer guidance.

The programming tasks were very diverse in their difficulty. We started
with the creation of the synthetic graphs discussed earlier in the
lecture. To make the task more feasible we gave the code to create
a path graph $P_{n}$ of $n$ nodes (see Listing~\ref{code:path}).
After a discussion of the code and answering questions like `Why is
the FOR loop progressing from $0$ to $n-1$?' they had to adopt the
code to create functions that construct other graphs such as the empty
graph $N_{n}$, the cycle graph $C_{n}$, and the complete graph $K_{n}$.
We registered that a subset of the pupils had problems with this task
and a careful discussion of the code was necessary. A line-by-line
dismantling of the script on the blackboard was especially helpful to increase
understanding.

Other tasks included the illustration of the networks with the \textsc{networkX}
{\tt draw} function and the computation of network metrics such as degree
and betweenness. For each exercise we also had one or two rather complex
tasks, such as the implementation of the Dijkstra-algorithm to keep
even the most advanced participants engaged. We clearly communicated
that we did not expect the pupils to solve all available tasks to
ensure they did not feel insufficient. We discussed the findings of
all tasks with the whole class.

\begin{lstlisting}[label={code:path},caption={Python code for creating a
        path graph $P_{10}$ with tehn nodes as provided to the students as starting
        point for further
        programming, as the creation of a complete graph $C_{8}$ with eight nodes.},language={Python},label={code:path}]
# import necessary library
import numpy as np

# define a function 
def path_graph( n ):

 #  Returns the (n x n) adjacency matrix 
 #  of a path graph with n nodes.


    # empty adjacency matrix of shape (n x n)
    A = np.zeros((n,n))

    # loop over every node besides the last one
    for u in range(0, n-1):

        # connect this node to the subsequent one
        A[u,u+1] = 1

    # symmetrize matrix by adding its transpose
    # the original matrix
    # this ensures that that not only u is conn-
    # ected to v, but the other way around, too.
    A = A + np.transpose(A)

    # the matrix will now be returned
    return A

# call the function to obtain an adjacency matrix
# for a path graph of 10 nodes
path_10 = path_graph(10)

# put the matrix on screen in text
print(path_10)
\end{lstlisting}

\subsection{Rotation}

\label{subsec:rotation}

The \emph{Rotation} is an inherent part of each DSA academy. It occurs
at half-way through the course, thus after approximately seven days of
covered material. It is a half-day event in which each course has
to organize twenty minute presentations for pupils of other courses.

We planned the talks in our course and discussed which topics were
appropriate to present to the other pupils. Our participants identified
that the fundamental mathematical definitions should be covered and
social networks would be a good example to make the topic accessible. They therefore covered the small-world phenomenon and the friendship
paradox.

The feedback they received on the presentations was very positive.
The participants as well as the lecturers of the other courses were astonished
by how applicable and approachable mathematics can be. We think that
network science itself can serve as a `figurehead' topic inside mathematics
because it easily contradicts the usual stereotypes of mathematics
being incomprehensible and overly complicated.

\subsection{Dokumentation}

\label{subsec:doku}

The \emph{Dokumentation} (German for documentation) is another inherent
part of each DSA summer school. It is a written report in which each course
summarizes the covered material. Each participant contributed a small
one to two page part, while we served as editors.

Despite most participants having no knowledge of \LaTeX{} we decided
to write the whole report with \textsc{ShareLaTeX}, an online \LaTeX{} editor
that allows real-time collaboration and online compiling~\cite{oswaldsharelatex}.\footnote{There are plenty of other \LaTeX{}\ online editors available, such as Overleaf,
Papeeria, and Authorea.} This came with the known advantages of \LaTeX{} such as  easy incorporation
of formulas, figures, and citations. It furthermore allowed the collaboration
of all pupils and removed compatibility issues. We introduced the
use of \LaTeX{}\ in a special session and accompanied the whole
writing process to efficiently dismantle encountered problems.

The first step was the discussion of an overall structure of the Dokumentation.
We did not impose any approach but discussed different potential approaches with the
pupils. The final structure was similar but not identical to the one
we chose for the course overall. The second step was the individual
discussion with each pupil about the material they wanted to cover
and what illustrations were necessary. All figures in this paper were
created by pupils as part of the documentation.

The creation of the Dokumentation was a very novel challenge for the
participants. One encountered problem was the space limitation, such
that the presented information had to be condensed, while staying
comprehensible. Furthermore, the appropriate referencing was unfamiliar
to the pupils. Most of them did not encounter these aspects before
and thus found them challenging. Overall however, we were impressed
with the result and needed only two to three iterations of editing
per pupil before the texts met the appropriate quality.

\section{Conclusions}

\label{sec:conclusions}

We were overall content with the course, the pupils' involvement,
and their progress. We think that network science is a appropriate
topic to give pupils an introduction to university-level science and
mathematics, as we noted from the feedback during the Rotation, even
for those that are not naturally interested in such topics.

We encountered the disparity in background, especially concerning programming,
as the most challenging part, because it made it more difficult to
keep everybody engaged. As two instructors, we were able to
split the class into a beginner and advanced half, which helped
in some cases. A better, more detailed, discussion of programming
for those that had never programmed before might have been fruitful.

A crucial part of our teaching was the flexibility in the covered
topics. Sometimes the pupils had problems with concepts that we thought
were taught at school, e.g.~the binomial coefficient, while other
new concepts took much less time to discuss than we anticipated.

One weakness of our preparation was the reader. We should have tailored
it more to the specific course content and given `homework' problems
instead of reading material only. This would have allowed us to challenge
the pupils with more advanced topics, such as more complicated differential
equations, and thus go into more detail in the course itself.

Although the discussion of a broad range of topics was praised by
the pupils we felt that a focus on networks under omission of a majority
of the dynamic systems part would have been beneficial. The latter
is an interesting and complex branch of science and would provide
enough material for a course by itself. In retrospect, we would definitely
drop the discussion of Lotka-Volterra systems, Turing patterns, chaos,
and fractals while a discussion of stability and infection dynamics
might still be interesting as an example for the application of network
science to real-world problems.

We also want to note that the organization of such a course was also very
beneficial for ourselves. It allowed us to recapture many different
topics in graph theory and network science and think about ways to
motivate and connect them with each other.

Finally, as well as \cite[Harrington \emph{at al.}]{harrington2013commentary} and \cite[Sanchez \emph{et al.}]{sanchez2014more},
we encourage fellow network scientists to use the presented material
to tailor their own course to spread the fascination about networks
and dynamic systems amongst teenagers, and to contact us in case there
are any questions. 

\section{Acknowledgements}

We would like to thank all the pupils who participated in the course for
their engagement and criticism. It was a great pleasure working
with them for the two weeks of the program. Further thanks goes to
the central organization of the DSA in Bonn, as well as the academy
administration and fellow course organizers at the DSA Ro\ss{}leben 2016.
We thank our doctoral advisers
Mason A. Porter and Dirk Brockmann for supporting us in this endeavor
and for their fruitful discussion. We are grateful to Sarah M. Griffin for proofreading the manuscript.

\section{Funding}
This work is supported by the EPSRC and MRC (FK, grant numbers EP/L016044/1, EP/R513295/1, EP/N014529/1) and by the Joachim Herz Stiftung (BFM).

\bibliographystyle{unsrt}
\bibliography{dsa_article.bib}

\end{document}